\documentclass[twocolumn,amsmath,amssymb,prl,showpacs]{revtex4}
\usepackage{graphicx,color}
\usepackage{dcolumn}
\usepackage{bm}
\newcounter{one}
\usepackage{braket}
\usepackage[tight]{subfigure}
\begin{document}

\title{Universal Quantum Computing with Measurement-Induced Continuous-Variable\\ Gate Sequence in a Loop-Based Architecture}

\author{Shuntaro Takeda}
\email{takeda@alice.t.u-tokyo.ac.jp}
\author{Akira Furusawa}
\email{akiraf@ap.t.u-tokyo.ac.jp}
\affiliation{Department of Applied Physics, School of Engineering, The University of Tokyo,\\ 7-3-1 Hongo, Bunkyo-ku, Tokyo 113-8656, Japan}

\date{\today}


\begin{abstract}
We propose a scalable scheme for optical quantum computing using measurement-induced continuous-variable quantum gates in a loop-based architecture. 
Here, time-bin-encoded quantum information in a single spatial mode is deterministically processed in a nested loop by an electrically programmable gate sequence.
This architecture can process any input state and an arbitrary number of modes with almost minimum resources,
and offers a universal gate set for both qubits and continuous variables.
Furthermore, quantum computing can be performed fault-tolerantly
by a known scheme for encoding a qubit in an infinite dimensional Hilbert space of a single light mode.
\end{abstract}



\maketitle


{\it Introduction.}---Quantum optical systems provide a promising platform to realize universal quantum computing.
Recently, there has been significant progress in the implementation of optical quantum logic gates by a measurement-induced scheme,
i.e., by using off-line prepared ancillary states, measurement, and feedforward~\cite{99Gottesman,03Bartlett,04Sanaka,11Okamoto,05Filip,07Yoshikawa,08Yoshikawa,14Miwa,11Marek,16Miyata}.
This scheme utilizes specific ancillae to perform quantum gates
which are difficult to perform directly on quantum states.
For continuous-variables (CVs), this scheme has offered deterministic gates
by means of unconditionally prepared squeezed ancillae and highly efficient homodyne detection~\cite{05Filip,07Yoshikawa,08Yoshikawa,14Miwa}.
This is a great advantage over qubits,
where ancilla preparation and photon detection are less efficient and
measurement-induced gates become probabilistic~\cite{04Sanaka,11Okamoto}.
For this reason, there has been a growing interest in a hybrid approach
combining robust encoding of qubits and deterministic gates of CVs~\cite{13Takeda,11Furusawa}.
This approach potentially enables scalable, universal, and fault tolerant quantum computing,
which is hard to achieve by either qubit or CV scheme alone~\cite{01Gottesman}.

Universal quantum computation can be realized by the sequence of such measurement-induced gates.
However, the previous implementation of such gates~\cite{04Sanaka,11Okamoto, 07Yoshikawa,08Yoshikawa,14Miwa}
encoded quantum information in spatial modes,
requiring a large number of optical components and resources for sequential gates.
For scalable quantum computing, integrated waveguides~\cite{15Carolan, 16Harris} have been developed to miniaturize optical circuits, but it still requires arrays of sources operating in parallel~\cite{17Spring}
and the stabilization of a large number of interferometers.
An alternative solution to the scalability is to employ time-bin encoding instead of spatial-mode encoding.
For CVs, time-bin encoding has been used for generating large-scale entangled states in a scalable manner~\cite{13Yokoyama,16Yoshikawa}.
As for qubits, linear optical quantum computing using time-bin encoding~\cite{13Humphreys}
and a loop-based architecture for processing time-bin encoded information~\cite{14Motes,15Rohde,12Schreiber,17He}
have been shown to be a useful platform for scalable quantum computing.

\begin{figure}[!b]
\vspace{-2mm}
\begin{center}
\includegraphics[width=1\linewidth,clip]{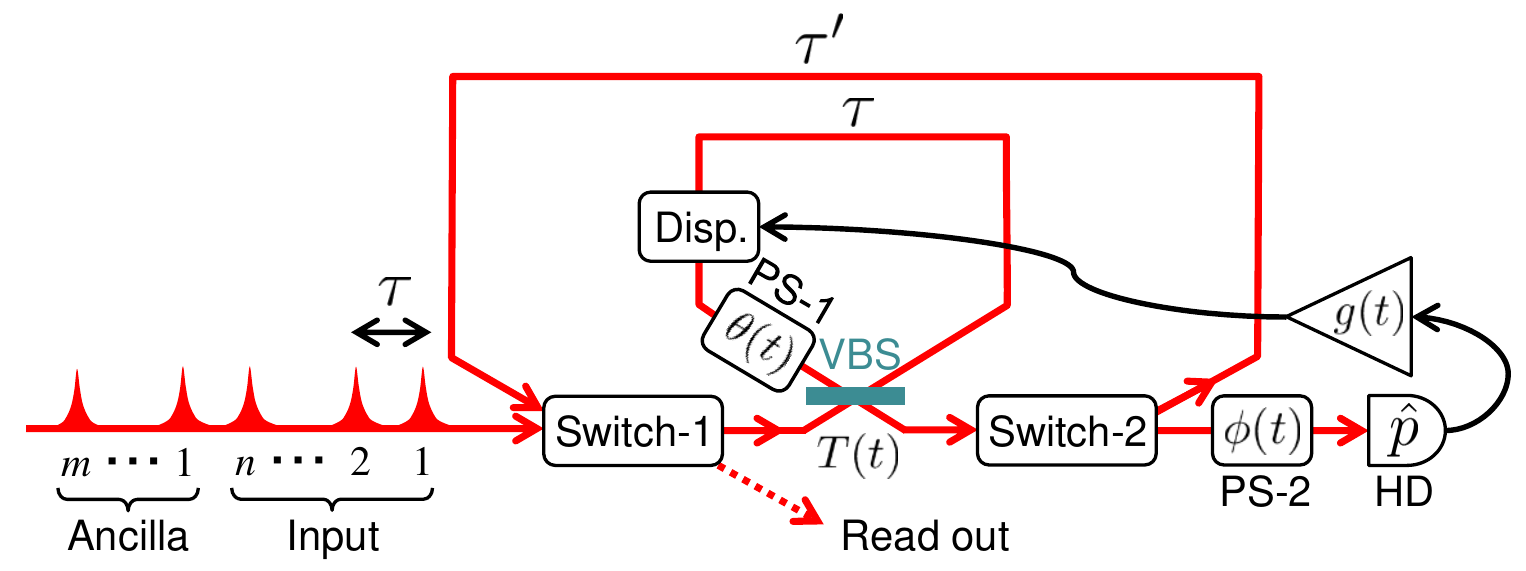}
\end{center}
\vspace{-5mm}
\caption{
Loop-based architecture for universal quantum computing.
HD, Homodyne Detector; Disp., Displacement operation;
PS, Phase shifter; VBS, Variable Beam Splitter.
}
\label{fig:NestedLoop}
\end{figure}

Here we propose a scalable scheme for quantum computing using measurement-induced CV gates in a loop-based architecture.
In our scheme (Fig.~\ref{fig:NestedLoop}),
quantum information encoded in a string of $n$ pulses of a single spatial mode are sent to a nested loop circuit with the other $m$ ancilla pulses.
Quantum gate sequence is then deterministically performed in the loop circuit with assistance of the ancilla pulses
by electrically programmable control of switches, beam splitter transmissivity, phase shifters, and amplifier gain.
Our scheme has several distinct features.
First, it can deal with any input state and an arbitrary number of modes in the same experimental complexity,
offering higher scalability than the conventional spatial-mode encoding schemes.
Second, our scheme is resource efficient:
it requires only one ancillary squeezing resource and a single set of a homodyne detector and feedforward electronics 
to deterministically perform an arbitrary multi-mode Clifford gates.
Finally, if suitable ancillae are provided,
it offers a universal gate set for both qubits and CVs.
Our scheme is fully compatible with a known error-correction scheme
which encodes a qubit in an infinite dimensional Hilbert space of a single light mode~\cite{01Gottesman},
ultimately enabling fault-tolerant quantum computation with almost minimum resources.

\begin{figure}[!b]
\vspace{-5mm}
\begin{center}
\includegraphics[width=1\linewidth,clip]{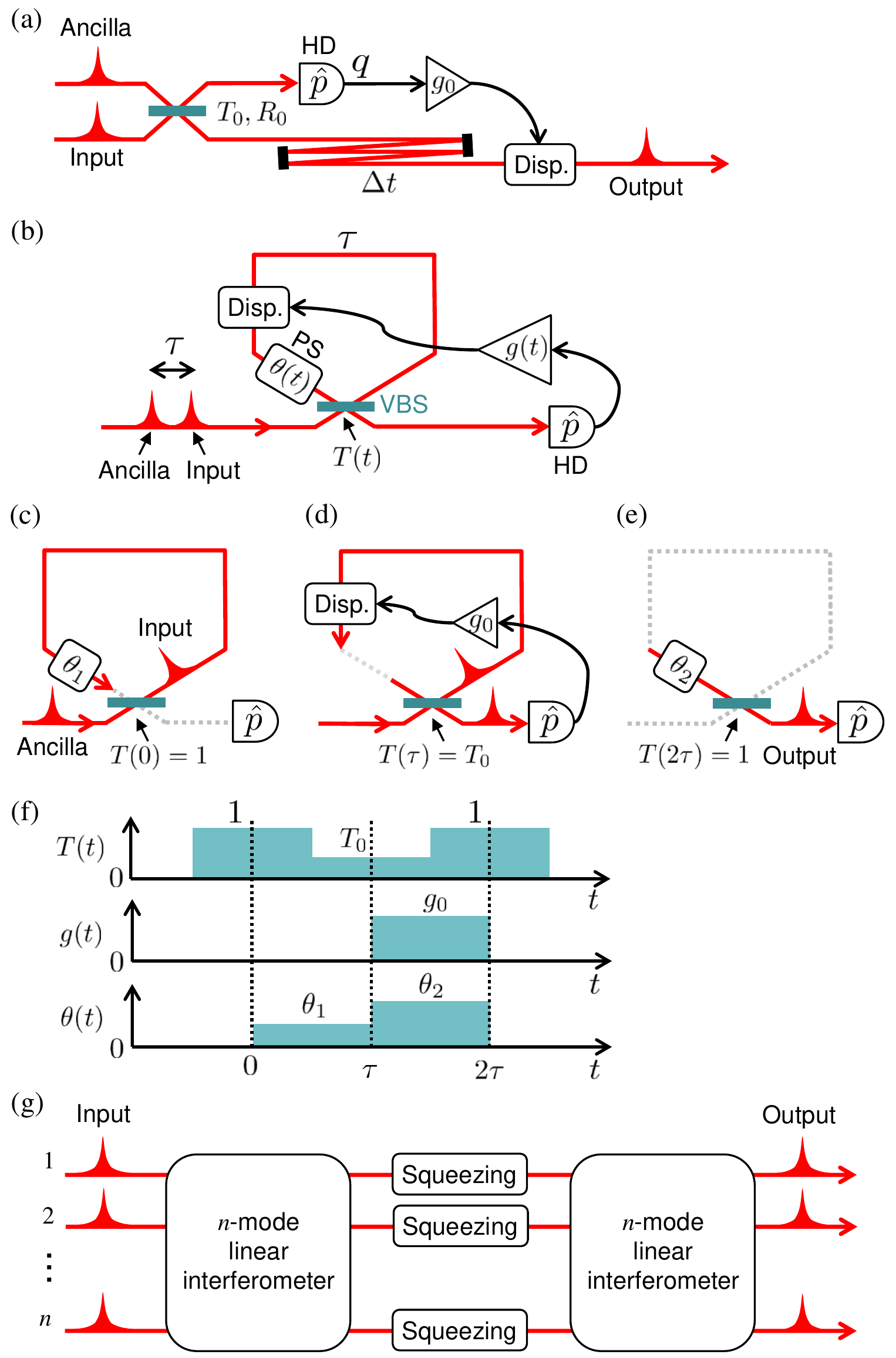}
\end{center}
\vspace{-2mm}
\caption{
(a) Measurement-induced squeezing gate~\cite{05Filip}.
(b) Single-loop architecture for single-mode Clifford (Gaussian) gates.
(c-e) Procedure for an arbitrary single-mode Clifford gate.
We assume that the input pulse initially arrives at VBS at time $t=0$.
(f) Programmable control sequence of system parameters.
(g) Decomposition of an arbitrary $n$-mode Clifford gate~\cite{05Braunstein}.
}
\label{fig:Gaussian}
\end{figure}

{\it Single-mode Clifford gates.}---Before describing the nested loop architecture in Fig.~\ref{fig:NestedLoop},
we begin by introducing a single loop architecture to implement an arbitrary single-mode Clifford gate.
In terms of CV operations, Clifford gates are equivalent to Gaussian gates,
which can be decomposed into the form $\hat{R}(\theta_2)\hat{S}(r)\hat{R}(\theta_1)$
up to a phase-space displacement~\cite{05Braunstein}.
Here $\hat{R}(\theta)$ is a phase-space rotation by angle $\theta$ and
$\hat{S}(r)$ $(r>0)$ is an $\hat{x}$-squeezing operator defined as
$\hat{S}^\dagger(r)\hat{x}\hat{S}(r)=e^{-r}\hat{x}$ and
$\hat{S}^\dagger(r)\hat{p}\hat{S}(r)=e^{r}\hat{p}$,
where $\hat{x}$ and $\hat{p}$ are quadrature operators of a light mode ($[\hat{x}, \hat{p}]=i$).
Since phase-space displacements and rotations can be easily implemented with optical modulators and phase shifters,
the main difficulty in single-mode Gaussian gates is the squeezing gate requiring second order nonlinear optical effects.
Instead of directly coupling fragile quantum states to nonlinear optical media for squeezing,
a measurement-induced squeezing scheme has been previously proposed~\cite{05Filip} and experimentally demonstrated~\cite{07Yoshikawa,14Miwa}.
The schematic is shown in Fig~\ref{fig:Gaussian}(a).
This scheme uses an ancillary squeezed state as a resource of nonlinearity.
First, an arbitrary input state is combined with an ancillary $\hat{x}$-squeezed vacuum state at a beam splitter with transmissivity $T_0$ and reflectively $R_0$ ($T_0+R_0=1$).
$\hat{p}$ quadrature of one of the output beams is measured by a homodyne detector.
Finally, the measurement outcome $q$ is fedforward to displace $\hat{p}$ quadrature of the other mode by $g_0q$, where $g_0=\sqrt{T_0/R_0}$ is an amplifier gain.
In the ideal limit of infinitely squeezed ancilla, input-output relation of this operation becomes
$\hat{x}_\text{out}=\sqrt{R_0}\hat{x}_\text{in}$ and
$\hat{p}_\text{out}=\hat{p}_\text{in}/\sqrt{R_0}$ in the Heisenberg picture,
where the subscripts ``out'' and ``in'' refer to the output and input modes, respectively~\cite{05Filip}.
These relations correspond to the squeezing operation $\hat{S}(-\ln\sqrt{R_0})$,
and the degree of squeezing can be controlled by choosing the beam splitter reflectivity $R_0$ and accordingly changing the gain $g_0$.
In the actual experiment of squeezing pulsed input states~\cite{14Miwa}, an optical delay line of time $\Delta t$ has to be introduced to
compensate the electric delay during the feedforward operation, as depicted in Fig~\ref{fig:Gaussian}(a).

The original circuit in Fig~\ref{fig:Gaussian}(a) encodes quantum states in spatial modes.
Now we convert this circuit into a single loop circuit in Fig~\ref{fig:Gaussian}(b) for time-bin encoding.
In Fig.~\ref{fig:Gaussian}(b), we assume that beam splitter transmissivity $T(t)$, phase shift $\theta(t)$, and gain $g(t)$ can be dynamically controlled.
Below we show that, by programmably controlling these system parameters,
an arbitrary single-mode Gaussian gate in the form $\hat{R}(\theta_2)\hat{S}(-\ln\sqrt{R_0})\hat{R}(\theta_1)$ can be performed in the loop circuit.
Suppose that input and ancillary squeezed pulses are sent to the loop circuit with time separation of $\tau$ [Fig.~\ref{fig:Gaussian}(b)].
First, the input pulse is picked up into the loop by setting the beam splitter transmissivity to $T(t=0)=1$ [Fig.~\ref{fig:Gaussian}(c)].
The pulse entering the loop takes time $\tau$ to circle around the loop,
and is subjected to a phase shift $\hat{R}(\theta_1)$ before coming back to the beam splitter again.
Then a measurement-induced squeezing gate is implemented in the loop circuit [Fig.~\ref{fig:Gaussian}(d)].
After one cycle in the loop, the input pulse coincides with the ancillary squeezed pulse at the beam splitter of $T(\tau)=T_0$.
The pulse leaving the loop is then immediately measured by a homodyne detector,
whose output signal is fedforward with gain $g_0$ to the pulse inside the loop.
The electric delay $\Delta t$ during the feedforward operation can be compensated
as long as the loop length $\tau$ is longer than $\Delta t$.
This feedforward completes the squeezing operation $\hat{S}(-\ln\sqrt R_0)$.
Finally the pulse in the loop is subjected to another phase shift $\hat{R}(\theta_2)$
and exits the loop by choosing $T(2\tau)=1$ [Fig~\ref{fig:Gaussian}(e)].
The whole control sequence of the system parameters is summarized in Fig~\ref{fig:Gaussian}(f),
and electrical programming of the sequence enables an arbitrary single-mode Gaussian gate
in the same experimental configuration.
If needed, displacement operations can be added just by sending desired signals to
the modulator instead of the homodyne detector's signal,
though it is not explicitly shown in Fig~\ref{fig:Gaussian}(b).
Therefore, this loop circuit provides a sufficient set of operations for an arbitrary single-mode Clifford gate
in a programmable fashion.
The loop plays a role of an optical delay enabling a beam splitter operation between different time-bins
as well as compensating the electric delay in the feedforward operation.
Since ancillary squeezed states can be continuously and deterministically supplied from an optical parametric oscillator,
all these operations can be performed deterministically with only one squeezing resource and a single set of a homodyne detector and feedforward electronics.

{\it Multi-mode Clifford gates.}---Next we extend the above scheme from single-mode to multi-mode Clifford gates.
As shown in Fig.~\ref{fig:Gaussian}(g), it is known that any $n$-mode Clifford (Gaussian) gate in CVs can be decomposed into
a $n$-mode linear interferometer, followed by the parallel applications of a set of single-mode squeezing operations,
followed by another $n$-mode linear interferometer~\cite{05Braunstein} (up to displacement operations).
Since the single-mode squeezing operation is already available in the loop circuit in Fig~\ref{fig:Gaussian}(b),
an arbitrary $n$-mode linear interferometer for time-bin encoding has to be introduced somehow.

An arbitrary $n$-mode linear interferometer can be decomposed into a sequence of  at most $n(n-1)/2$ pairwise beam splitter operations~\cite{94Reck}.
Such a sequence of beam splitter operations can be implemented
by embedding the loop circuit in Fig~\ref{fig:Gaussian}(b) into a larger loop of round-trip time $\tau^\prime$~\cite{14Motes, 15Rohde}, as shown in Fig.~\ref{fig:NestedLoop}.
The larger loop is controlled by another two optical switches.
When a string of $n+m$ pulses ($n$ input and $m$ ancillary squeezed pulses, $m\le n$) are sent to the circuit, ``Switch-1'' lets all the pulses enter the outer loop.
This requires $\tau^\prime\ge(n+m)\tau$.
While these pulses repeatedly circulate the outer loop,
$n$ input pulses pass through the inner loop to implement a sequence of pairwise beam splitter operations.
The programmable control sequence of switches, beam splitter transmissivity $T(t)$, and phase shifter $\theta(t)$
enables the implementation of an arbitrary $n$-mode beam splitter operation.

Once the initial $n$-mode interferometer in Fig.~\ref{fig:Gaussian}(g) is implemented,
$m$ ancillary squeezed pulses are used to sequentially implement measurement-induced squeezing gates
by controlling ``Switch-2'' and activating the feedforward operation (phase shifter 2 is not used at the moment).
Finally another $n$-mode interferometer is implemented to complete the desired Gaussian gate.
Note that the pulses after the operation can be read out
from the unused output port of ``Switch-1,'' as shown in Fig.~\ref{fig:NestedLoop}.
As a result, the loop architecture in Fig.~\ref{fig:NestedLoop}
enables an arbitrary multi-mode Clifford gate deterministically
based on a programmable control sequence.
By increasing the outer loop length $\tau^\prime$, our architecture can deal with an arbitrary number of modes in the same experimental complexity.

{\it A non-Clifford gate.}---In addition to the multi-mode Clifford gates described above,
at least one non-Clifford, or non-Gaussian gate is necessary and sufficient to complete the CV universal gate set~\cite{99Lloyd}.
Non-Gaussian gates require third or higher order nonlinear effects, such as Kerr effect, which are hard to implement directly on quantum states.
In contrast, a measurement-induced scheme for one of the non-Gaussian gates, a cubic phase gate, was proposed in Refs.~\cite{11Marek,16Miyata,01Gottesman}.
In this scheme, cubic nonlinearity can be deterministically supplied with an ancillary state which can be prepared probabilistically~\cite{13Yukawa,16Ogawa} and stored in quantum memories~\cite{13Yoshikawa}.
Below we show that the cubic phase gate proposed in Ref.~\cite{16Miyata} can be implemented in the architecture of Fig.~\ref{fig:NestedLoop}.

\begin{figure}[!b]
\vspace{-2mm}
\begin{center}
\includegraphics[width=1\linewidth,clip]{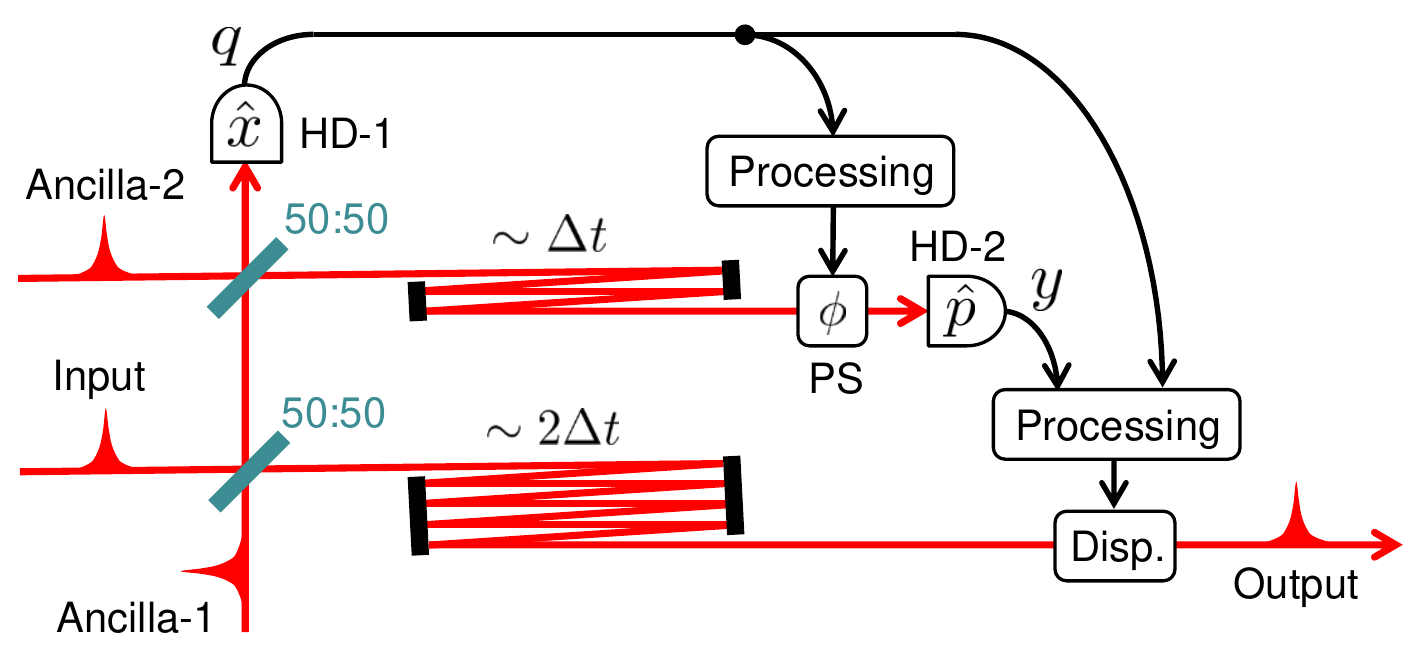}
\end{center}
\vspace{-5mm}
\caption{
Measurement-induced cubic phase gate in Ref.~\cite{16Miyata}.
}
\label{fig:NonGaussian}
\end{figure}

A cubic phase gate $\hat{C}(\gamma)$ is defined as
$\hat{C}^\dagger(\gamma)\hat{x}\hat{C}(\gamma)=\hat{x}$
$\hat{C}^\dagger(\gamma)\hat{p}\hat{C}(\gamma)=\hat{p}+3\gamma\hat{x}^{2}$.
Figure~\ref{fig:NonGaussian} shows the implementation of this gate in Ref.~\cite{16Miyata}.
The input state is first combined with an ancillary $\hat{x}$-squeezed vacuum state (Ancilla-1) at a $50:50$ beam splitter.
One of the combined modes is further combined with another ancillary state $\int dx \exp(i\gamma^\prime x^3)\ket{x}$ (Ancilla-2) at a $50:50$ beam splitter,
and the two resultant modes are measured by homodyne detectors HD-1 and HD-2.
Before HD-2, the phase of the pulse needs to be shifted by $\phi=\arctan(3\sqrt{2}\gamma^\prime q)$,
where $q$ is the outcome of $\hat{x}$ measurement at HD-1.
Finally, $\hat{p}$ quadrature of the remaining mode is displaced by $\sqrt{2}y/\cos\phi$, where $y$ is the outcome of $\hat{p}$ measurement at HD-2.
In the limit of infinite squeezing and perfect ancillary states,
the input-output relation is given in the Heisenberg picture by~\cite{16Miyata}
\begin{align}
\hat{x}_\text{out}=\frac{\hat{x}_\text{in}}{\sqrt{2}},  \;\;
\hat{p}_\text{out}=\sqrt{2}\left(\hat{p}_\text{in}+\frac{3\gamma^\prime}{2\sqrt{2}}\hat{x}_\text{in}^2\right),
\end{align}
and corresponds to $\hat{S}(\ln\sqrt2)\hat{C}(\gamma^\prime/2\sqrt2)$.
The unnecessary squeezing gate can be canceled out by applying the corresponding anti-squeezing operation. 

Once the input and two ancilla pulses are sequentially fed to our architecture of Fig.~\ref{fig:NestedLoop} ($n=1$ and $m=2$),
the same gate can be straightforwardly implemented by 
performing beam splitter operations, measurement, and feedforward in an appropriate order.
We have to note that more complicated control of the system parameters is necessary compared to the previous Gaussian gates.
First, the phase shifter $\phi$ before HD-2 in Fig.~\ref{fig:NonGaussian}
is implemented by phase shifter 2 in Fig.~\ref{fig:NestedLoop},
and the shifted amount nonlinearly depends on the measurement outcome of homodyne detection.
In addition, the displacement in the last step requires a classical signal
which nonlinearly depends on two measurement outcomes obtained at different time.
These kinds of complicated control are, however, implementable by appropriately rewriting the program.
Therefore, a cubic phase gate can be deterministically implemented in our architecture
when necessary ancillae are injected into the circuit.
This non-Gaussian cubic phase gate, together with already available multi-mode Gaussian gates,
constitute the universal gate set for CV quantum computation.

{\it Hybrid approach.}---All the above Clifford and non-Clifford gates can be applied to any input state, let alone qubits.
Thus,  by taking a hybrid approach combining qubit encoding and CV operations,
our architecture offers scalable and universal quantum computation for qubits.
For example, our architecture is compatible with a time-bin qubit, which is a superposition of a single photon in either of two pulses:
$\ket{\psi}=\alpha\ket{1,0}+\beta\ket{0,1}$.
For this encoding, any single-qubit operation can be directly performed in our architecture with only beam splitter operations on the two pulses.
Universal quantum computation also requires a two-mode entangling gate, which can be realized with Kerr interaction~\cite{00Nielsen}.
This interaction is also implementable in our architecture
by an appropriate sequence of only Gaussian operations and cubic phase gates~\cite{11Sefi}.

Furthermore, the hybrid approach enables fault-tolerant quantum computation with almost minimum resources
by redundantly encoding a qubit in a large Hilbert space of a single temporal mode.
One important example in Ref.~\cite{01Gottesman}
encodes a qubit in the logical basis
$\ket{j_\text{L}}=\sum_{s\in\mathbb{Z}}\ket{x=\sqrt{\pi}(2s+j)}$ ($j=0, 1$).
This qubit can be protected against sufficiently small phase-space displacement errors
by ancilla-assisted quantum error correction (Fig.~4 in Ref.~\cite{01Gottesman}).
Except for ancilla preparation, all the requirements for this error correction protocol,
including quantum non-demolition interaction (two-mode Gaussian operation) between a qubit and an ancilla,
and homodyne measurement followed by a displacement operation,
are already included in our architecture.
Note that fault tolerance in this scheme is achievable even with non-ideal
qubit and CV cluster states with finite squeezing~\cite{14Menicucci}.
This conclusion also holds true for our architecture
since it can create CV cluster states from ancillary squeezed states~\cite{07vanLoock} in the loop and follow the same error-correction protocol in principle.
The threshold value of squeezing given in Ref.~\cite{14Menicucci} is 20.5 dB.
This means that losses in the loop of our architecture must be at least below $1\%$ to
keep the sufficient level of squeezing for fault tolerance.
These requirements are still higher than state-of-the-art technology,
considering the fact that the current highest squeezing level is 15 dB~\cite{16Vahlbruch}.
However, the requirement for fault tolerance is likely to be satisfied in the near future
by further improvement of technology or error-correction protocols. 
More comprehensive analysis of fault tolerance, including the effect of finite squeezing and losses in the loop,
can be performed with the extension of Ref.~\cite{14Menicucci} and will be given in a future work.

{\it Experimental viability.}---Finally, let us consider the feasibility of our architecture with current technology.
Measurement-induced squeezing gates in our architecture
was previously demonstrated in the configuration of Fig.~\ref{fig:Gaussian}(a) for pulsed input states~\cite{14Miwa}.
Although the measurement-induced cubic phase gate in Fig.~\ref{fig:NonGaussian} has not been demonstrated yet,
recent progress in ancilla preparation~\cite{13Yukawa,16Ogawa}
makes its implementation within reach of current technology.
Conversion of these gates into the loop architecture in Fig.~\ref{fig:NestedLoop}
requires long ($\gtrsim10$ m) and low-loss optical delay lines.
Such delay lines have been developed in free-space or by using optical fibers
in several experiments~\cite{13Yokoyama, 15Kaneda, 13LIGO}.
In addition, the viability of loop-based beam splitter operation has been recently demonstrated
in several experiments, such as quantum walk~\cite{12Schreiber} and boson sampling~\cite{17He}.
These experiments demonstrate that fast dynamic control
of optical switches and beam splitter transmissivity is possible while preserving the coherence of quantum states.
The demonstrations described above show that all of the basic building blocks of our architecture are already available.

{\it Conclusion.}---In conclusion, we showed that our quantum computation scheme using
measurement-induced CV gates in a loop-based architecture
provides the universal gate set for both qubits and CVs.
This architecture offers electrical programmability of gate sequence and higher scalability,
and also enables fault-tolerant quantum computation with logical qubits redundantly encoded in a large Hilbert space.


\begin{acknowledgments}
This work was partly supported by CREST (JPMJCR15N5) of JST, JSPS KAKENHI, APSA.
S. T. acknowledges Hisashi Ogawa for his useful comments on the manuscript.
\end{acknowledgments}


\end{document}